\documentclass[prl,twocolumn,showpacs,preprintnumbers,aps,nofootinbib]{revtex4-2}

\usepackage{graphicx}
\usepackage{dcolumn}
\usepackage{bm}
\usepackage{amsmath,amssymb,amsfonts}
\usepackage{latexsym}

\usepackage{color}

\def\nn    {\nonumber}

\def\rbb{\rho_{bb}}
\def\rtt{\rho_{tt}}

\begin{document}


\title{\boldmath
Probing Baryogenesis with Radiative Beauty Decay and Electron EDM}

\author{Wei-Shu Hou,$^{1}$ Girish Kumar$^1$ and Tanmoy Modak$^2$}
\affiliation{$^1$Department of Physics, National Taiwan University, Taipei 10617, Taiwan}
%
%
\affiliation{$^2$Institut f\" ur Theoretische Physik, Universit\"at Heidelberg, 69120 Heidelberg, Germany }

 \date{\today}

\begin{abstract} 
With the Large Hadron Collider (LHC) running, we should probe electroweak 
baryogenesis (EWBG) while probing $CP$ violation (CPV) with electron 
electric dipole moment (eEDM). Rooted in the flavor structure of the 
Standard Model (SM), the general two Higgs doublet model (g2HDM) 
with a second set of Yukawa couplings can deliver EWBG while surviving 
eEDM. We point out a chiral-enhanced top-bottom interference effect that 
makes $b \to s\gamma$ decay an exquisite window on EWBG and eEDM, 
and illustrate the importance of the $\Delta A_{CP}$ observable at Belle~II.
\end{abstract}

\maketitle


\noindent{\it Introduction.---}
The rare $b \to s\gamma$ process offers 
the best bound~\cite{Misiak:2017bgg} on the charged Higgs boson $H^+$ that 
exists in two Higgs doublet models~\cite{Branco:2011iw}, in particular 
2HDM-II that is automatic with supersymmetry (SUSY). It holds true even 
after 15 years of LHC running, and a decade since the 
discovery~\cite{h125_discovery} of  $h(125)$, the 125 GeV boson. 

The $h$ boson completes one Higgs doublet, 
so extra Higgs bosons from a second doublet {\it must} be searched for. 
2HDM-II is one of two models that obey 
the natural flavor conservation (NFC) condition of Glashow and 
Weinberg~\cite{Glashow:1976nt}: each type of fermion charge couples 
to {\it only} one Higgs doublet; in 2HDM-II, $u$- and $d$-type quarks 
couple to different doublets. But since {\sl No New Physics}, SUSY 
included, has emerged at the LHC, one should see the NFC condition 
as it is: {\it ad hoc}. 

In this Letter we study the general 2HDM, i.e. the {\it natural} case 
of having two Yukawa matrices. The mass matrix is diagonalized as usual, 
giving $\lambda_f = \sqrt2\, m_f/v$ for $f = u, d, \ell$ with $v$ the 
vacuum expectation value (VEV); it has been confirmed~\cite{PDG} 
for $t$, $b$, $\tau$ and $\mu$ at the LHC. 
The second Yukawa matrix, $\rho^f$, cannot be simultaneously diagonalized 
in general, hence 
the fear of flavor changing neutral couplings (FCNC).
But as shown long ago~\cite{Cheng:1987rs},  taking some fermion 
mass-mixing Ansatz that reflects the observed hierarchical pattern, NFC 
may not be needed. It was stressed that mass-mixing hierarchies alone 
may be Nature's way to control FCNC~\cite{Hou:1991un}, 
with $t \to ch$ the likely harbinger, which has been 
pursued~\cite{PDG} ever since the $h(125)$ discovery. The current limit 
of 0.073\%~\cite{CMS:2021hug} is getting stringent.

We promote g2HDM as a likely {\it next} New Physics. 
Most important is its ability to deliver~\cite{Fuyuto:2017ewj} EWBG~\cite{EWBG}: the disappearance of antimatter shortly after the Big Bang, i.e.~the 
Baryon Asymmetry of the Universe (BAU). 
First, g2HDM with ${\cal O}(1)$ Higgs quartic couplings~\cite{Kanemura:2004ch}
 can give first order phase transition. 
Second, complex $\rho^f$ couplings can give large CPV 
with three mechanisms: most robust is via $\rho_{tt}$ at ${\cal O}(\lambda_t)$, 
i.e. ${\cal O}(1)$~\cite{Fuyuto:2017ewj}, and with $|\rho_{tc}| \simeq 1$ 
as back up if  $\rho_{tt}$ turns out accidentally small; $\rho_{bb}$ can also 
give~\cite{Modak:2018csw, Modak:2020uyq} EWBG if its strength is large 
enough, but would need $10^{-3}$ tuning.

The emergent {\it alignment} phenomenon, that $h$ resembles the SM Higgs 
boson~\cite{PDG} so well, means that the $h$--$H$ mixing angle 
$c_\gamma \equiv \cos\gamma$ 
is small, where $H$ is the exotic $CP$-even scalar. As the SM Higgs 
boson cannot induce $t \to ch$ decay, the coupling is 
$\rho_{tc}c_\gamma$~\cite{Chen:2013qta}, i.e.~$\rho_{tc}$ relates to 
the exotic doublet that has no VEV. With the stringent 
$t \to ch$~\cite{CMS:2021hug} bound, Nature seems to throw in a 
non-flavor, purely Higgs-sector parameter $c_\gamma$ to help 
suppress the $t \to ch$ FCNC process.

Alignment is not~\cite{Hou:2017hiw} in conflict with the need of ${\cal O}(1)$ 
Higgs quartics, e.g. the $h$--$H$ mixing coupling $\eta_6 \sim 1$~\cite{Hou:2017hiw} 
is allowed even for relatively small $c_\gamma$. 
Further, EWBG implies exotic $H$, $A$ and $H^+$ bosons should be 
sub-TeV in mass --- a boon to LHC search~\cite{Kohda:2017fkn,
 Ghosh:2019exx, Hou:2020chc}.

The large CPV in g2HDM for EWBG does bring on a {\it general challenge}: 
surviving eEDM constraints, which has recently leapfrogged neutron EDM 
(nEDM). 
The impressive ACME bound~\cite{ACME:2018yjb} 
was recently surpassed by JILA, to
 $0.41 \times 10^{-29}\,e$\,cm~\cite{Roussy:2022cmp}. 
However, a ``natural" {\it flavor-based} cancellation can 
evade~\cite{Fuyuto:2019svr} eEDM bounds elegantly. 
%
To cancel quite a few two-loop effects due to $\rho_{tt}$ and {\it also}
$\rho_{ee}$, one needs 
\begin{align}
   |\rho_{ee}/\rho_{tt}| = r|\lambda_e/\lambda_t|,
   \ \ \ \ \arg(\rho_{ee}\rho_{tt}) = 0,
\label{cancel}
\end{align}
with $r \sim 0.7$, depending on loop functions. The second ``phase-lock'' 
neutralizes the effect from pseudoscalar $A$~\cite{Fuyuto:2019svr}, giving the 
first relation that implies the $\rho^f$ matrices {\it know} about flavor hierarchies.
%

Considering how well g2HDM {\it evades} flavor constraints, 
we conceived~\cite{Hou:2020itz} a rule of thumb:
\begin{align}
   \rho_{ii} \lesssim {\cal O}(\lambda_i),\;\;
   \rho_{1i} \lesssim {\cal O}(\lambda_1),\;\;
   \rho_{3j} \lesssim {\cal O}(\lambda_3),
\label{rho_ij}
\end{align}
for $j \neq 1$, giving $\rtt = {\cal O}(1)$ but $\rbb \simeq 0.02$.
Indeed, $|\rtt| \lesssim 0.6$ is allowed~\cite{Ghosh:2019exx} by 
LHC data, and can further soften by finite $\rho_{tc}$. But we do 
not really know the parameter values. Since $\rtt$, $\rbb$ can 
each induce EWBG, in view of the prowess of eEDM --- discovery 
may come around $10^{-30}\,e$\,cm! --- we point out 
an $m_t/m_b$ chiral-enhanced $H^+$ effect of $\rtt$--$\rbb$ interference 
that compensates the smallness of $\rbb$ to make $b \to s\gamma$ an 
exquisite probe of EWBG. 
%
%
We explore next-to-leading order (NLO) effects for 
future development.

\vskip0.08cm
\noindent{\it Formalism.---}
We assume $CP$-conserving~\cite{Hou:2017hiw, Davidson:2005cw} 
Higgs potential of g2HDM, removing it as a CPV source to simplify, 
without discussing it further.
To clarify the flavor discussion given in the Introduction, the 
Yukawa couplings to charged fermions~\cite{Davidson:2005cw,
 Hou:2020chc} are
\begin{align}
\mathcal{L} = 
 - & \frac{1}{\sqrt{2}} \sum_{f = u, d, \ell}
 \bar f_{i} \Big[\big(-\lambda^f_i \delta_{ij} s_\gamma + \rho^f_{ij} c_\gamma\big) h \nn\\
  & + \big(\lambda^f_i \delta_{ij} c_\gamma + \rho^f_{ij} s_\gamma\big)H
    - i\,{\rm sgn}(Q_f) \rho^f_{ij} A\Big]  R\, f_{j} \nn\\
 - & \bar{u}_i\left[(V\rho^d)_{ij} R-(\rho^{u\dagger}V)_{ij} L\right]d_j H^+ \nn\\
 - & \bar{\nu}_i\rho^L_{ij} R \, \ell_j H^+ + {\rm h.c.},
\label{eff}
\end{align}
with family indices $i$,\,$j$ summed over,\;$L,\,R =\,1\,\mp\,\gamma_5$,\,and 
$s_\gamma \equiv \sin\gamma$. The $A$,\,$H^+$ couplings do not depend on 
$c_\gamma$; in the alignment limit 
($c_\gamma\,\to\,0$, $s_\gamma\,\to\,-1$),\,$h$\,couples diagonally and $H$ couples 
via\,$-\rho_{ij}^f$. Thus, besides mass-mixing hierarchy 
protection\,\cite{Hou:1991un}\,of\,FCNC, alignment gives\,\cite{Hou:2017hiw} further safeguard, 
without the need of NFC.

We follow the $b \to s\gamma$ formalism of Ref.~\cite{Ciuchini:1997xe}
 (see also Ref.~\cite{Borzumati:1998tg}). 
By replacing $A_u\to \rho_{tt}/\lambda_t$, $A_d\to \rho_{bb}/\lambda_b$, 
one-loop corrections to the Wilson coefficients (WCs) $C_7$ and $C_8$ 
induced by $H^+$ in g2HDM are, 
\begin{align}
       \delta C_{7, 8}^{(0)}(\mu)
  =  \frac{|\rho_{tt}|^2}{3|\lambda_t|^2} F_{7, 8}^{(1)}(x_t)
    - \frac{\rho_{tt}\rho_{bb}}{\lambda_t\lambda_b}F_{7, 8}^{(2)}(x_t),
\label{eq:C78}
\end{align}
with $x_t=m_t(\mu)^2/m_{H^+}^2$ at heavy scale $\mu$, and loop functions 
$F_{7, 8}^{(i)}(x)$ ($i=1, 2$) given in Ref.~\cite{Ciuchini:1997xe}. From 
$B_d$ and $B_s$ mixing constraints~\cite{Altunkaynak:2015twa}, we set 
$\rho_{ct} = 0$ .
%

%
While the form of Eq.\;\eqref{eq:C78} is correct, 
the\,denominators, i.e.\;the\;$|\lambda_t|^2$\;and\;$|\lambda_t\lambda_b|$\;factors\;actually arise from balancing explicit {\it masses} for $H^+$\;couplings in 2HDM-I\;\&\;II, 
rather than from dynamical couplings as the numerators. 
Thus, the second $\rtt$-$\rbb$ interference term receives 
$m_t/m_b$ chiral enhancement. 
Unlike chiral enhancement in left-right symmetric 
models~\cite{Fujikawa:1993zu, Cho:1993zb}, in Eq.~\eqref{eq:C78} 
it is rooted in the chiral $H^+$ couplings of Eq.~\eqref{eff}, 
where its origins will be elucidated further later.

As $\rtt$ and $\rbb$ can each lead to EWBG, we define 
\begin{align}
       \phi \equiv \arg(\rtt\rbb) = \phi_{tt} +  \phi_{bb},
\label{phi}
\end{align}
%
and illustrate with $\phi = 0,\,\pi$, $\pm\pi/2$, as explained later.

At NLO in QCD, $\delta C_{7, 8}$ at scale $\mu$ are defined as
\begin{align}
       \delta C_{7, 8}(\mu)
  =  C_{7, 8}^{(0)}(\mu) + \frac{\alpha_s (\mu)}{4\pi} C_{7, 8}^{(1)}(\mu),
\label{eq:C78_nlo}
\end{align}
where $\delta C_{7, 8}^{(0)}$ are given in Eq.~\eqref{eq:C78}, and 
$\delta C_{7, 8}^{(1)}$ are 
\begin{align}
       \delta C_{7}^{(1)}(\mu)
 & = G_{7}(x_t) + \Delta_7(x_t)\log \frac{\mu^2}{m_{H^+}^2} - \frac{4}{9}E(x_t), \\
       \delta C_{8}^{(1)}(\mu)
 & = G_{8}(x_t) + \Delta_8(x_t)\log \frac{\mu^2}{m_{H^+}^2} - \frac{1}{6}E(x_t),
\end{align}
with\;$G_{7, 8}(x)$,\;$\Delta_{7, 8}(x)$ and\;$E(x)$ given in 
Ref.~\cite{Ciuchini:1997xe}.
For 2HDM-I\;and\;II results at NNLO, see Ref.\;\cite{Hermann:2012fc}.

\vskip0.08cm
%
\noindent{\it Numerical results.---}
We consider the following $b \to   s\gamma$\;\cite{Paul:2016urs} observables: 
inclusive $B \to X_s \gamma$; exclusive\;$B^{+,\,0} \to K^\ast \gamma$ and 
$B_s \to \phi \gamma$, and CP asymmetries $A_{\rm CP}(B^{+,\,0} 
\to K^{\ast} \gamma)$; and the inclusive CPV difference~\cite{Benzke:2010tq} 
$\Delta A_{\rm CP}(b\to s\gamma) \equiv A_{\rm CP}(B^+ \to X_s^+ \gamma) 
- A_{\rm CP}(B^0 \to X_s^0 \gamma)$.
More observables 
can be included, but do not improve the bounds.

We illustrate with $m_{H^+} = 300$, $500$ GeV. 
For exclusive modes, $B \to V$ ($V = K^\ast, \phi$) form factors are needed. 
We follow Ref.~\cite{Paul:2016urs} and use the package
 \texttt{Flavio}~\cite{Straub:2018kue} for our estimation. 
The WCs in Eq.~\eqref{eq:C78} at heavy scale $\mu \sim m_{H^+}$ should 
be evolved down to the physical scale~\cite{Straub:2018kue} of 2 ($4.8$) GeV 
for inclusive (exclusive) processes, which is done using the package
 \texttt{Wilson}~\cite{Aebischer:2018bkb}.

\begin{figure*}[t!]
\center
\includegraphics[width=0.38 \textwidth]{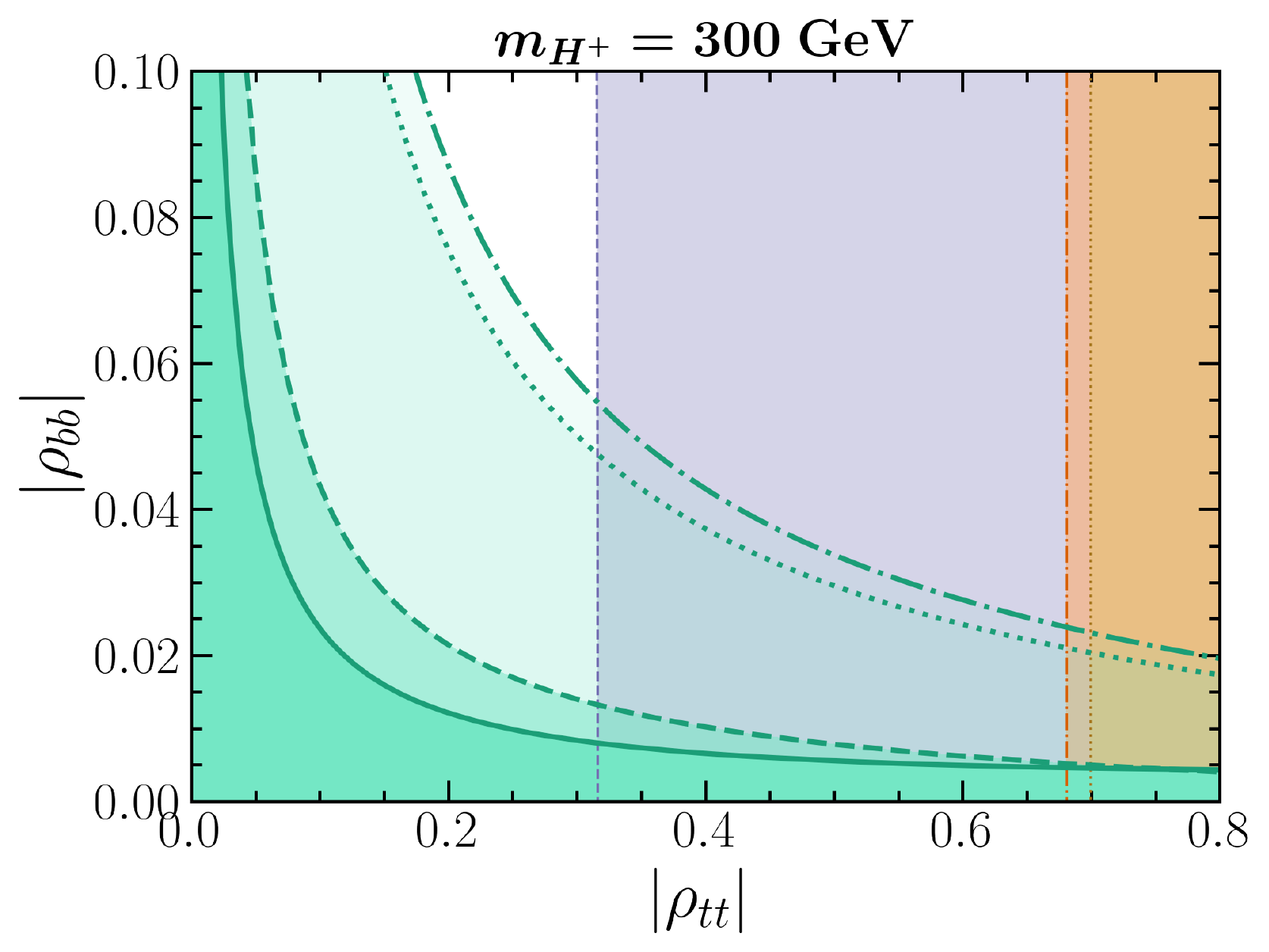}
\includegraphics[width=0.38 \textwidth]{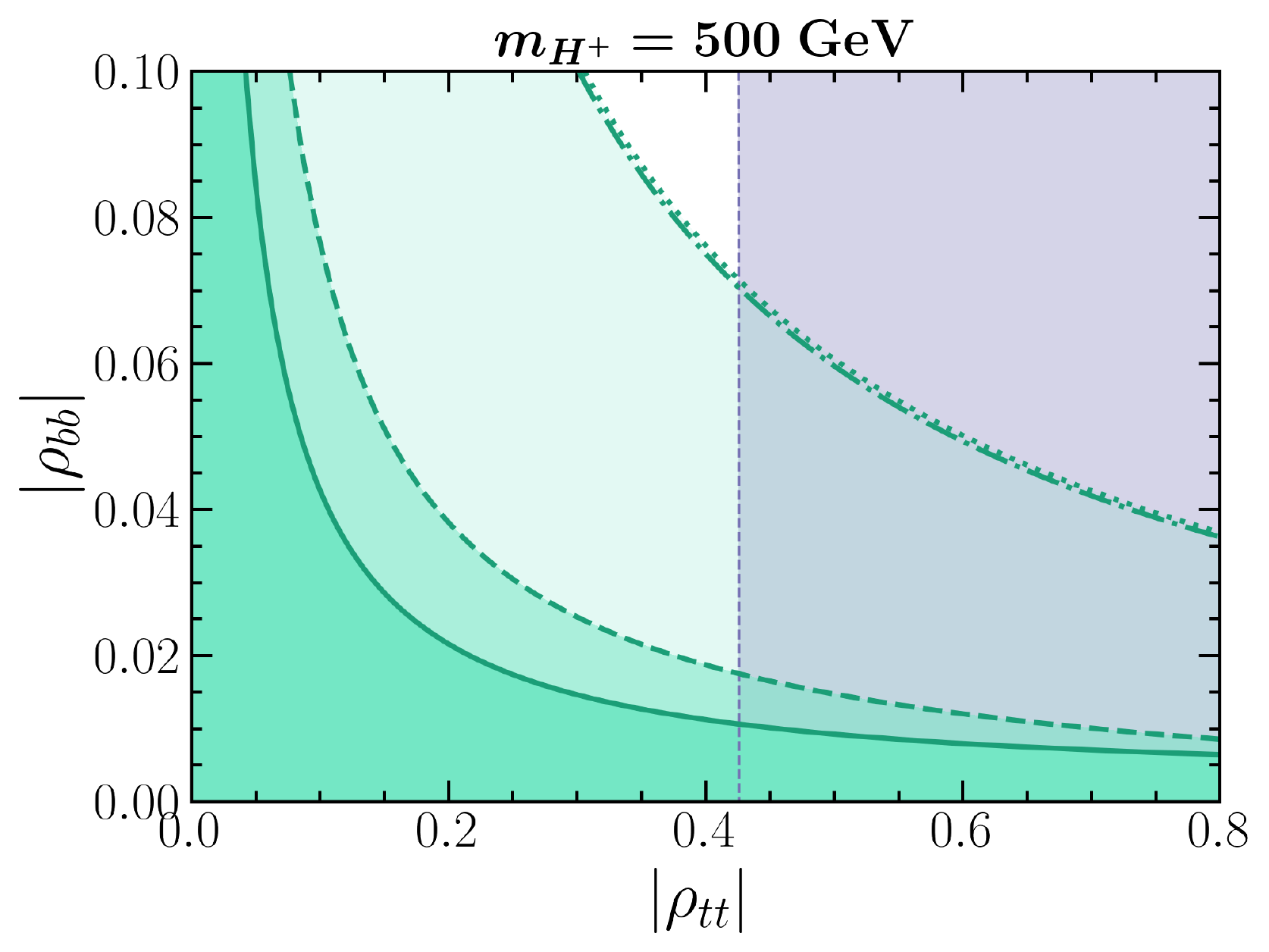}
\caption{
 {\sl Allowed regions} 
 from $b\to s\gamma$ observables, given for
 $\phi = \phi_{tt} + \phi_{bb} = 0$ (green solid), $\pi$ (green dashed),
 $+\pi/2$ (green dot-dash), {and $-\pi/2$} (green dots) for
 $m_{H^+} = 300$  (left), 500 (right) GeV.
 Constraints from $\Delta m_{B_s}$ (light blue short-dash),
 $B_s \to \mu\mu$ (red dot-dash) and $\varepsilon_K$ (brown dots)
 on $\rtt$ are also shown, with shaded region ruled out to the right.
}
\label{fig:rtt-rbb}
\end{figure*}

\begin{figure*}[t!]
\center
\includegraphics[width=0.38 \textwidth]{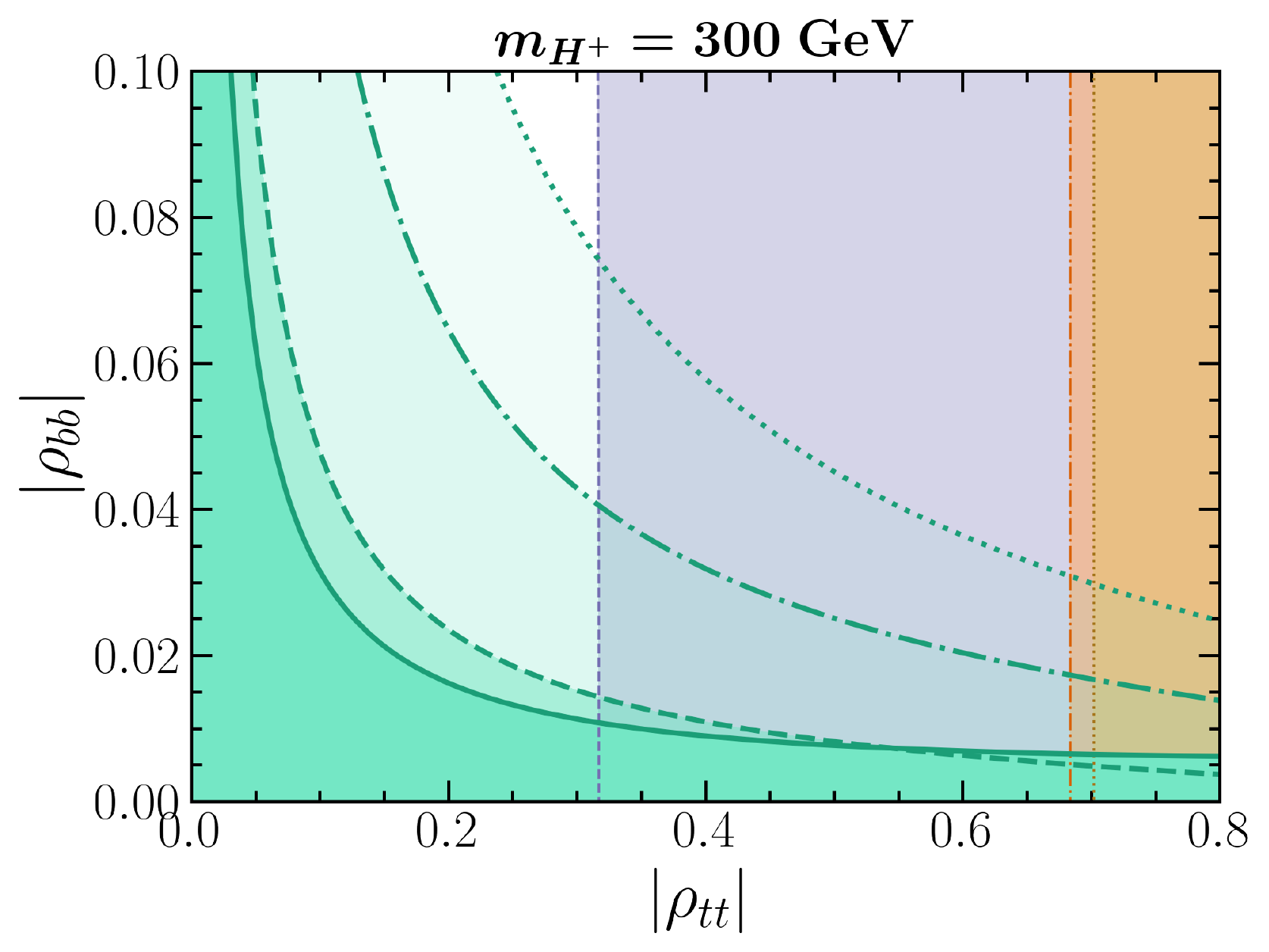}
\includegraphics[width=0.38 \textwidth]{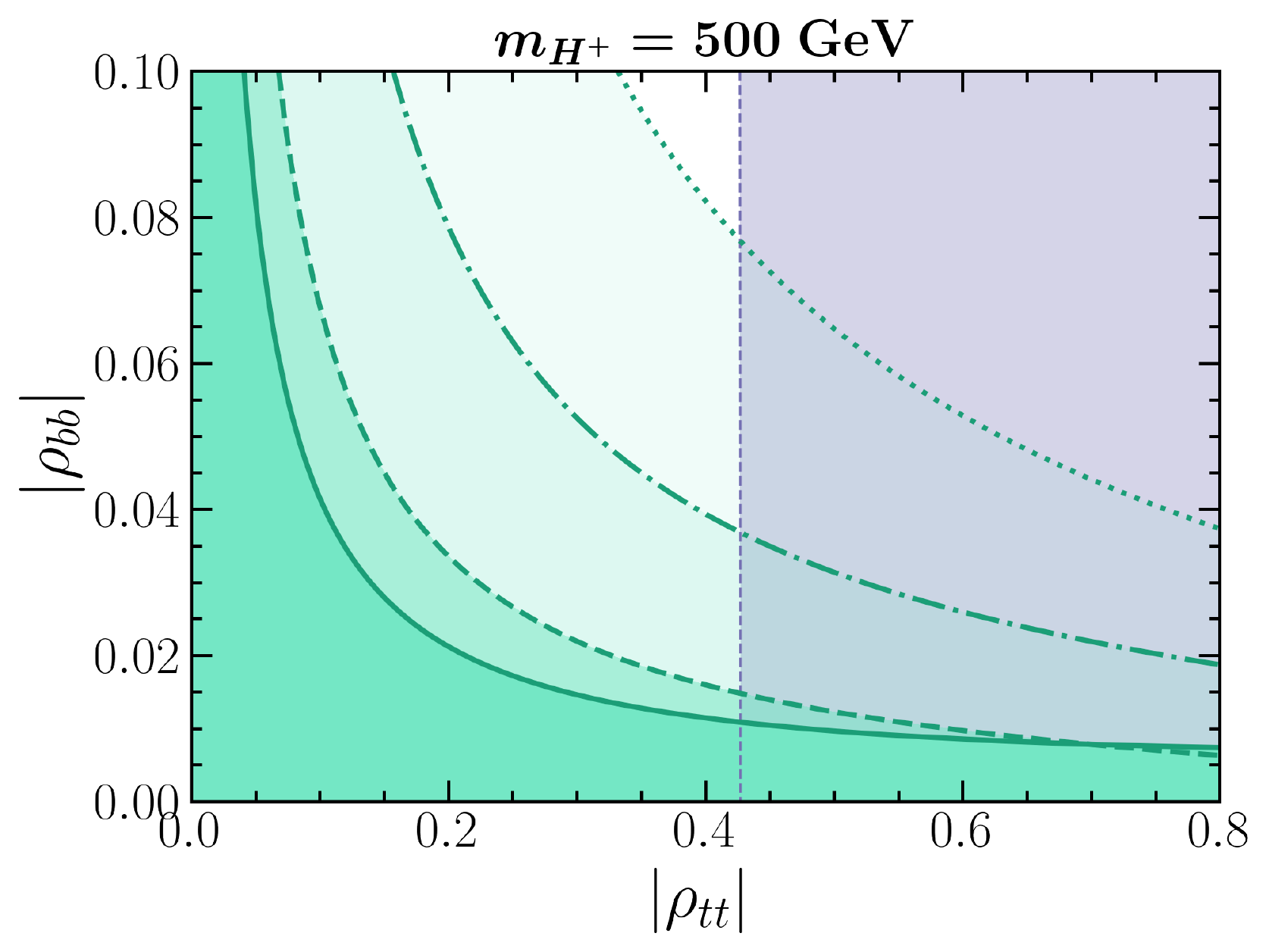}
\caption{
 Same as Fig.~\ref{fig:rtt-rbb}, but for $b \to s\gamma$ constraint at NLO.
}
\label{fig:rtt-rbb_nlo}
\end{figure*}

\texttt{Flavio v2.4.0} and experimental values for inclusive
 ${\cal B}(B \to X_s \gamma)|_{E_\gamma > 1.6 {\rm \,GeV}}$
branching ratios are:
\begin{align}
      & (3.29 \pm 0.22) \times 10^{-4}, \quad \  (\texttt{Flavio}\text{\;\cite{Straub:2018kue}})\\
      & (3.49 \pm 0.19) \times 10^{-4}, \quad \  (\text{HFLAV\;\cite{HFLAV:2022wzx}})
\end{align}
compared with the detailed theory value of 
${\cal B}(B \to X_s \gamma)|_{E_\gamma\,>\,1.6
 {\rm \,GeV}}\,=\,(3.40\,\pm\,0.17)\,\times\,10^{-4}$\,\cite{Misiak:2020vlo}, 
where error is closer to experiment. We use \texttt{Flavio} to cover more observables 
to compare with data~\cite{HFLAV:2022wzx} and show that, 
with $\rtt$ and $\rbb$ both present, $b \to s\gamma$ is 
the best probe of EWBG parameter space in g2HDM.

As only the combined phase $\phi = \phi_{tt} + \phi_{bb}$ of Eq.~\eqref{phi} 
enters, 
we plot in Fig.~\ref{fig:rtt-rbb} the global $b \to s\gamma$ constraint 
at leading order (LO) in $|\rtt|$--$|\rbb|$ plane, for $m_{H^+}=$ (left) 300 
and (right) 500\,GeV, illustrating for\;$\phi = 0, \pi, \pm\pi/2$, 
where shaded region is {\it allowed}, i.e. ruled out to the right. 
We also show three flavor constraints as vertical lines:\;$B_s$\;mixing\;\cite{Hou:2022qvx}, 
${\cal B}(B_s \to \mu\mu)$\;\cite{CMS:2022mgd}, 
and $\varepsilon_K$\;\cite{Hou:2022qvx}, 
where shaded regions to the right are ruled out.
Only $1\sigma$ bounds are shown to illustrate the prowess of 
$b\to s\gamma$ as probe, otherwise the two weaker flavor bounds 
would fly out of the plot. We comment on this later.

We see that, while $\Delta m_{B_s}$ limits $\rtt$ strength, thanks to 
chiral enhancement, the $b \to s\gamma$ constraint is more exquisite, 
probing even small $|\rbb| \lesssim 0.02$ values when $\rtt$ is sizable. 
The 
$\delta C_{7, 8}$ corrections (see Eq.~\eqref{eq:C78}) are small compared 
to the SM effect, which is enhanced by QCD~\cite{Bertolini:1986th,
 Deshpande:1987nr} and is close to real, which is why the $\phi = \pm\pi/2$ 
cases are more accommodating, as the $H^+$ effect sums only in quadrature.

We also give the result for NLO by taking $\delta C_{7, 8}^{(1)}$ of 
Eqs.~(7) and (8) into Eq.~\eqref{eq:C78_nlo}, run down from $\mu$
scale to low scale, and plot in Fig.~\ref{fig:rtt-rbb_nlo}, which is visibly 
different from Fig.~\ref{fig:rtt-rbb}. We leave to the experts for proper 
refinement.
Not shown are $m_{H^+} = 1$ TeV results, where the parameter space 
is more generous, as expected.

From Eq.~\eqref{rho_ij}, if we take $|\rbb| \sim 0.02$ to mean the range of 
$0.01 \lesssim |\rbb| \lesssim 0.03$, for $m_{H^+} = 300\; (500)$ GeV in 
Fig.~\ref{fig:rtt-rbb}. {For the most stringent $\phi = 0$ case, the bounds 
are $|\rtt| \lesssim0.08$ (0.14) for $\rho_{bb} = 0.03$,
and $|\rtt| \lesssim 0.24$ (0.43) for $\rho_{bb} = 0.01$.} 
{These $\rtt$ strengths are still more or less robust for 
EWBG~\cite{Fuyuto:2017ewj}, while $\rbb$ seems a bit small to be the driver.
However, 
if $\rtt$ turns out much less than 0.1 and ineffective for EWBG, we see from 
Fig.~\ref{fig:rtt-rbb} that $|\rbb| \sim 0.1$ becomes allowed by $b \to s\gamma$ 
and could~\cite{Modak:2018csw, Modak:2020uyq} drive EWBG.} 
For the $\rho_{tc}$ mechanism that evades eEDM, the $t\to ch$ 
bound~\cite{CMS:2021hug} puts some stress on $|\rho_{tc}| \sim 1$, 
despite alignment assistance. 
Note that 
Refs.~\cite{Fuyuto:2017ewj, Modak:2018csw, Modak:2020uyq} 
has the known issue of overestimating BAU compared to the semiclassical 
approach~\cite{Cline:2021dkf}, which is especially severe for the 
$\rho_{bb}$-EWBG mechanism. But due to large uncertainties in 
several parameters~\cite{Modak:2021vre}, it may still be open 
as one awaits a more precise estimation.

The $\phi = 0$ case may be special. In the $\rho_{ee}$--$\rtt$ eEDM 
cancellation mechanism, the second relation of Eq.~\eqref{cancel} 
imposes a phase-lock, that $\phi_{ee}$ is opposite in sign to $\phi_{tt}$. 
In estimating the CPV $e$-$N$ scattering correction, Ref.~\cite{Fuyuto:2019svr} 
took the ``Ansatz'' of $\phi_{qq} = - \phi_{tt}$. While not written in stone, 
we would like to bring in a ``bias'' from charge unification, that in context of 
grand unified theories (GUT), charged leptons and $d$-type quarks seem 
grouped together. Thus, while $u$-type quarks may not have this ``phase-lock'' 
with $\phi_{tt}$, $\phi_{bb} + \phi_{tt} = 0$ may be plausible, hence favor 
$\phi = 0$. 
{If no CPV is observed in $b \to s\gamma$ with all 
Belle~II~\cite{Belle-II:2018jsg} data, $\phi = 0$ (or $\pi$) may be suggested. 
Even so, the $\phi_{tt}$ and $\phi_{bb}$ phases can still contribute 
separately to EWBG.}

\begin{figure*}[t!]
\center
\includegraphics[width=0.38 \textwidth]{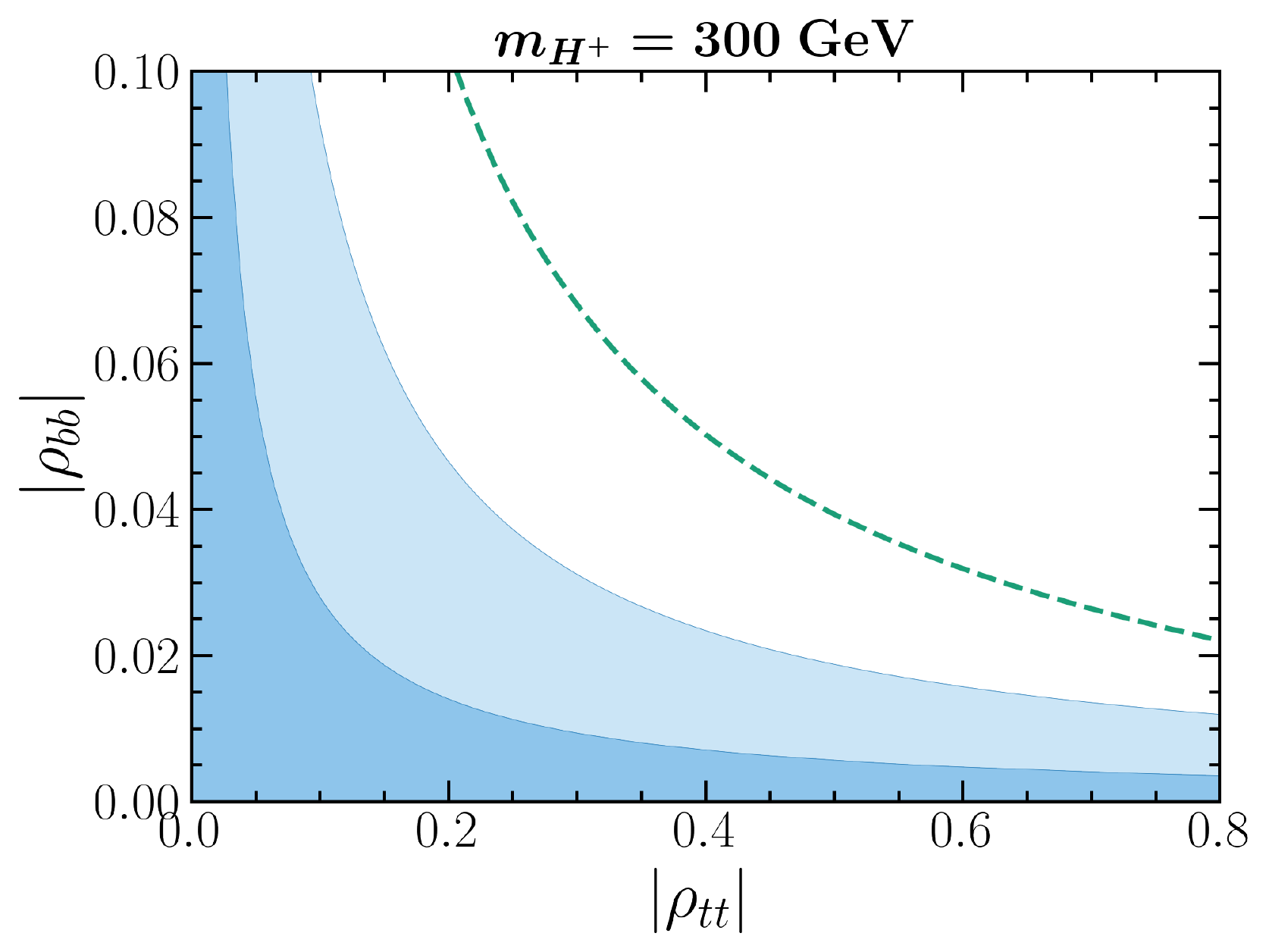}
\includegraphics[width=0.38 \textwidth]{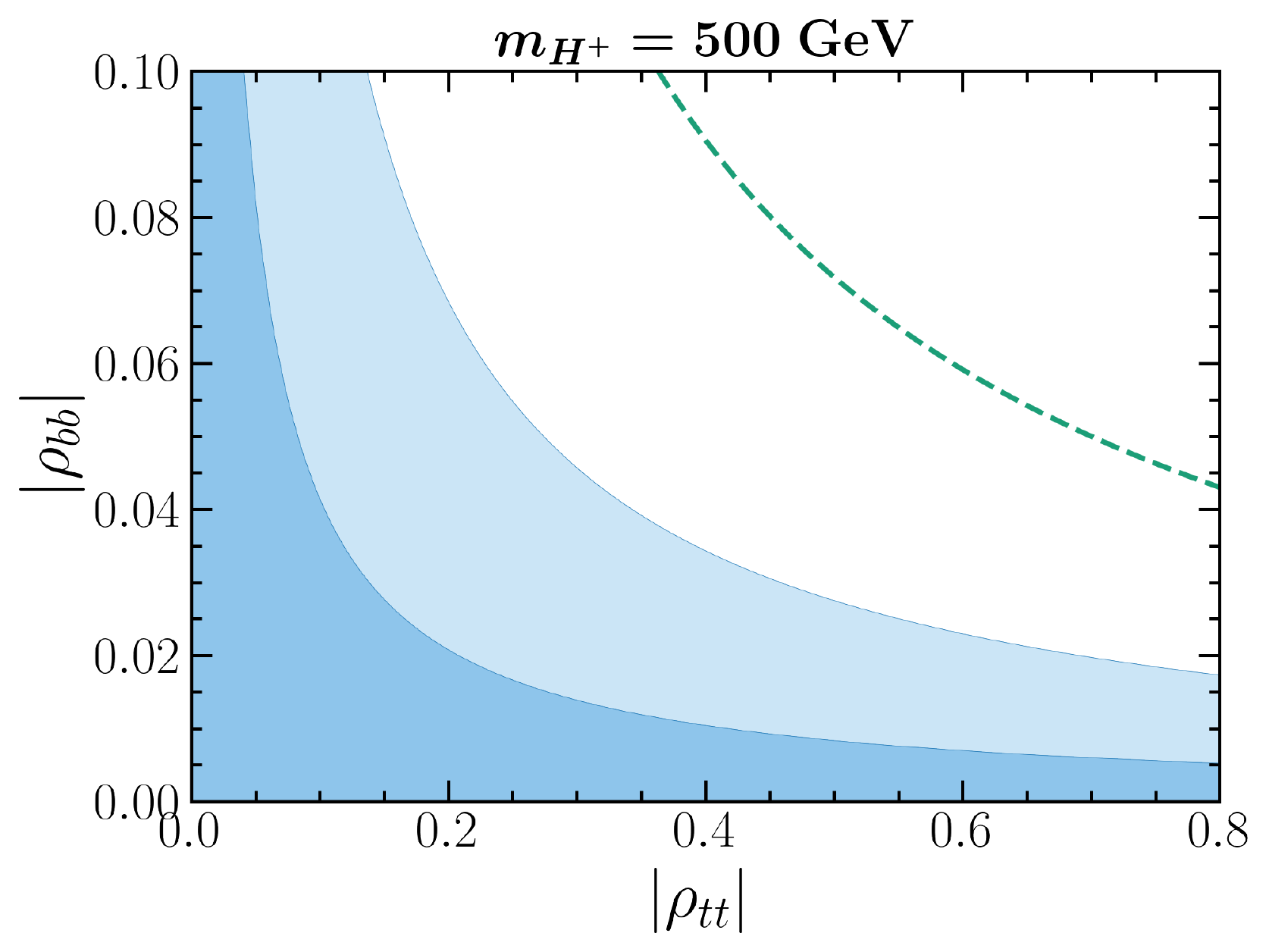}
\caption{
 (Light) blue shaded region {\it allowed} by $\Delta A_{\rm CP} (\simeq 0$
 assumed) with ($5\;{\rm ab}^{-1}$) $50\;{\rm ab}^{-1}$ Belle~II  data,
 and $\mathcal{B}(B \to X_s \gamma)$ constraint (green dot-dash) for
 $\phi=\pm \pi/2$. All are $1\sigma$ constraints.
}
\label{fig:rtb_cp_future}
\end{figure*}

\begin{table*}[t!]
\centering
  \begin{tabular}{lcccc}
  \toprule
  ~~$m_{H^+}$ & \multicolumn{3}{c@{}}{$\rho_{tt}$ benchmark} & \ \ $\rho_{bb}$
  benchmark \\
    & $|\rho_{tt}| = 0.1 \sqrt{2}$ & $|\rho_{tt}| = 0.2 \sqrt{2}$ & $|\rho_{tt}| = 0.3 \sqrt{2}$
    & $|\rho_{bb}| = 0.1$ \\
    \hline 
    $300$ GeV \ \ & $\mp (3.041 \pm 0.046)$ & $\mp (6.026 \pm 0.091)$
                          & $\mp (8.902 \pm 0.134)$ & \ \ $\mp (5.352\pm 0.080)$\\
     $500$ GeV \ \ & $\mp (2.055 \pm 0.031)$ & $\mp (4.097 \pm 0.063)$
                           & $\mp (6.111 \pm 0.093)$ & \ \ $\mp (3.628 \pm 0.055)$ \\
     \hline 
  \end{tabular}
  \caption{Expected $\Delta A_{\rm CP}$ ($\times 10^{-3}$) for
  {$\phi = \pm\pi/2$. In $\rho_{tt}$ ($\rho_{bb}$) benchmark,
  $|\rho_{bb}|=0.02$ ($|\rho_{tt}| = 0.05$) is used for illustration.}}
\end{table*}

Future Belle~II measurement of inclusive CPV difference 
$\Delta A_{\rm CP}$~\cite{Benzke:2010tq} 
can probe the phase $\phi$. 
{We show the 1$\sigma$ constraint on $\Delta A_{\rm CP}$ in 
Fig.~\ref{fig:rtb_cp_future} for 5 and $50\;{\rm ab}^{-1}$ data, 
for $\phi = \pm \pi/2$ and $m_{H^+} =$ (left) 300, (right) 500~GeV. 
The $\mathcal{B}(B \to X_s \gamma)$ bound is also shown, which does 
not improve by much: $\Delta A_{\rm CP}$ {\it indeed} probes $\phi$.}

{The eEDM cancellation mechanism was recently extended to broader
parameter range~\cite{Hou:2023kho}, and an nEDM cancellation mechanism 
was illustrated by variation of $\rho_{uu}$ strength and phase. We show in 
Table I that, interestingly, for the largest $\rho_{tt}$ strength and phase, 
$\Delta A_{\rm CP}$ could approach 3$\sigma$ with full Belle~II data. 
If eEDM emerges soon, and with good prospects for nEDM~\cite{Hou:2023kho}, 
one may face the decision on extending Belle~II data taking, 
or even upgrade to Belle~III.}

\vskip0.08cm
\noindent{\it Discussion and Conclusion.---}
Chiral enhancement was noted in Ref.~\cite{Altunkaynak:2015twa}, 
but the $\rho^{u,d}$ matrices were taken as real. 
Here, with large CPV from complex $\rtt$ and $\rbb$ contributing 
to EWBG and eEDM, we study specifically the effect of 
$\phi = \phi_{tt} + \phi_{bb}$ on $b \to s\gamma$.

So what is the origin of this chiral enhancement? Analogous to the elucidation
given in one~\cite{Hou:1987kf} of the earliest works on $b \to s\gamma$ in 
2HDM-I \& II, one needs a $\bar s\sigma_{\mu\nu} m_b Rb$ dipole structure, 
which could arise in two ways: 
from $H^+$ coupling to the internal top at both ends of the loop, thereby
 $\propto |\rtt|^2$, but would need an $m_b$ insertion in the external $b$ line; 
or $H^\pm$ couplings to $\rbb$ at $b$ quark end while $\rtt$ at $s$ quark end. 
To achieve the chirality flip in $\bar s\sigma_{\mu\nu}Rb$, an $m_t$ insertion 
is needed, resulting in the $m_t/m_b$ chiral enhancement.

Our figures 
show $1\sigma$ bounds to contrast with other flavor constraints. 
We could show $2\sigma$ constraints, but note that \texttt{Flavio} 
errors~\cite{Straub:2018kue} are 50\% larger than experiment~\cite{HFLAV:2022wzx}. 
As experimental errors improve at Belle~II~\cite{Belle-II:2018jsg}, 
theory needs to improve as well, 
which we expect~\cite{Belle-II:2018jsg} will happen.

With JILA surpassing the ACME to reach $0.41 \times 10^{-29}\,e$\,cm, 
the eminence of eEDM goes without saying. They are, however, still 
consistent with $10^{-29}\,e$\,cm. Given that g2HDM can achieve EWBG, with 
an exquisite cancellation mechanism for $\rtt$-EWBG while the less elegant 
$\rbb$-EWBG is also possible, a few $\times 10^{-30}\,e$\,cm would be quite 
contentious, as the {\it likelihood} within g2HDM is large. 
If eEDM emerges soon, 
it would provide support for EWBG \`a la g2HDM.
%
Any other EWBG proposal with large New Physics CPV would have
to pass the eEDM test.
As the cancellation mechanism for $\rtt$-EWBG  invokes flavor hierarchies, 
Eq.~\eqref{cancel}, while Nature seems to provide flavor protection against 
a plethora of probes~\cite{Hou:2020itz}, both seem to point to g2HDM: 
having a second Higgs doublets but without NFC condition.

Flavor physics does provide~\cite{Hou:2020itz} a set of probes, such as
${\cal B}(B \to \mu\nu)/{\cal B}(B \to \tau\nu)$~\cite{Hou:2019uxa} and 
$\tau \to \mu\gamma$ at Belle~II, $B_s \to \mu\mu$ at CMS and LHCb, 
$K^+ \to \pi^+\nu\nu$ at NA62 for heavier $H^+$~\cite{Hou:2022qvx}, 
and the possible revival of muon physics in $\mu \to e\gamma$,
 $\mu \to 3e$ and $\mu N \to eN$~\cite{Hou:2020itz}.
Direct search for the sub-TeV exotic $H$, $A$, $H^+$ Higgs bosons at the 
LHC~\cite{Kohda:2017fkn, Ghosh:2019exx, Hou:2020chc} should also be 
earnestly pursued, where ATLAS has made the first step~\cite{ATLAS:2023tlp}.

In conclusion,
g2HDM can provide electroweak baryogenesis while surviving electron 
EDM constraint, a remarkable feat that is rooted in the flavor structure 
as revealed in the SM sector. With exotic $H$, $A$ and $H^+$ bosons 
sub-TeV in mass, search programs at the LHC have started, while there 
are also some good flavor probes.
In this work we show that $b \to s\gamma$ offers an exquisite window on
baryogenesis and eEDM via a chiral enhancement of a special $t$-$b$
interference effect. With ongoing efforts at Belle~II and other flavor fronts,
and exotic Higgs search at the LHC, together with the supercharged eEDM 
front, the future looks bright for unveiling what may actually lie behind 
baryogenesis.

\vskip0.1cm
\noindent{\bf Acknowledgments} \
We thank the support of grants NSTC 111-2639-M-002-004-ASP, 
NTU 112L104019 and 112L893601. TM is supported by grants 
DFG 396021762-TRR 257 
and 
EXC 2181/1-390900948. 


\end{document}